\documentclass[aps,prl,reprint,nofootinbib,twocolumn,superscriptaddress,showpacs,showkeys,longbibliography]{revtex4-1}

\usepackage{amsmath,amssymb,amstext}
\usepackage{epsfig,epsf}
\usepackage{multirow}
\usepackage[usenames,dvipsnames]{color}
\usepackage{graphicx}
\usepackage{braket}
\usepackage{natbib}
\usepackage{comment}
\usepackage{dcolumn}
\usepackage[english]{babel}
\usepackage{wasysym}
\usepackage{bm}
\usepackage[colorlinks=true,linkcolor=blue,citecolor=blue,pdfauthor={ },pdftitle={ },pdfsubject={ },pdfkeywords={ }]{hyperref}
\usepackage{CJK}
\usepackage{lipsum}
\usepackage{color,url}

\graphicspath{{Figures/}}

\makeatother

\usepackage{array}      
\usepackage{amsmath}    
\usepackage{siunitx}    
\begin{document}
	
\title{Detection of missing low-lying atomic states in actinium}
\author{Ke Zhang}
\email{kezhang@uni-mainz.de}
\affiliation{Helmholtz Institute Mainz, GSI Helmholtzzentrum f\"{u}r Schwerionenforschung, 55099 Mainz, Germany}
\affiliation{Johannes
	Gutenberg-University Mainz, 55099 Mainz, Germany}
\author{Dominik Studer}
\affiliation{Johannes
	Gutenberg-University Mainz, 55099 Mainz, Germany}
\author{Felix Weber}
\affiliation{Johannes
	Gutenberg-University Mainz, 55099 Mainz, Germany}
\author{Vadim M. Gadelshin}
\affiliation{Johannes
	Gutenberg-University Mainz, 55099 Mainz, Germany}
\affiliation{Institute of Physics and Technology, Ural Federal University, 620002 Yekaterinburg, Russia}
\author{Nina Kneip}
\affiliation{Johannes
	Gutenberg-University Mainz, 55099 Mainz, Germany}
\author{Sebastian Raeder}
\affiliation{Helmholtz Institute Mainz, GSI Helmholtzzentrum f\"{u}r Schwerionenforschung, 55099 Mainz, Germany}
	\affiliation{Johannes
	Gutenberg-University Mainz, 55099 Mainz, Germany}
\author{Dmitry Budker}
\affiliation{Helmholtz Institute Mainz, GSI Helmholtzzentrum f\"{u}r Schwerionenforschung, 55099 Mainz, Germany}
	\affiliation{Johannes
	Gutenberg-University Mainz, 55099 Mainz, Germany}
\affiliation{Department of Physics, University of California at Berkeley, California 94720-300, USA}
\author{Klaus Wendt}
\affiliation{Johannes Gutenberg-University Mainz, 55099 Mainz, Germany}
\author{Tom Kieck}
\affiliation{Johannes Gutenberg-University Mainz, 55099 Mainz, Germany}
\affiliation{Helmholtz Institute Mainz, GSI Helmholtzzentrum f\"{u}r Schwerionenforschung, 55099 Mainz, Germany}
\author{Sergey G. Porsev}
\affiliation{Department of Physics and Astronomy, University of Delaware, DE, 19716, USA}
\affiliation{Petersburg Nuclear Physics Institute of NRC ``Kurchatov Institute'', Gatchina 188300, Russia}
\author{Charles Cheung}
\affiliation{Department of Physics and Astronomy, University of Delaware, DE, 19716, USA}
\author{Marianna S. Safronova}
\affiliation{Department of Physics and Astronomy, University of Delaware, DE, 19716, USA}
\affiliation{Joint Quantum Institute, NIST and the University of Maryland, College Park, MD, 20742, USA}
\author{Mikhail G. Kozlov}
\affiliation{Petersburg Nuclear Physics Institute of NRC ``Kurchatov Institute'', Gatchina 188300, Russia}
\affiliation{Petersburg Electrotechnical University ``LETI'', Prof. Popov Str. 5, 197376 St.~Petersburg, Russia}

\begin{abstract}

Two lowest-energy odd-parity atomic levels of actinium, $7s^27p\ ^{2}P^{o}_{1/2}$, $7s^27p\ ^{2}P^{o}_{3/2}$, were observed
via two-step resonant laser-ionization spectroscopy and their respective energies were measured to be
$7477.36(4)~\rm{cm}^{-1}$ and $12\,276.59(2)~\rm{cm}^{-1}$. The lifetimes of these states were determined as 668(11)\,ns
and 255(7)\,ns, respectively. In addition, these properties were calculated using a hybrid approach that combines configuration interaction and
coupled-cluster methods in good agreement. The data are of relevance for understanding
the complex atomic spectra of actinides and for developing efficient laser-cooling
and ionization schemes for actinium, with possible applications for high-purity medical-isotope production and future fundamental physics experiments with this atom.
\end{abstract}

\pacs{03.67.Lx,76.60.-k,03.65.Yz}
\maketitle

Actinium ($\rm{Z}=89$) lends its name to the actinide series, of which it is the first member.
The longest-lived isotope of actinium  $^{227}$Ac ($\tau_{1/2}\approx 22\,$y) is found in trace amounts as a member in the decay chain of natural $^{235}\rm{U}$. 
Actinium isotopes can be produced in nuclear reactors enabling their use in various applications based on their specific radioactivity. The isotope $^{225}\rm{Ac}$, an $\alpha$-emitter with a half-life of 10 days, is used in cancer radiotherapy \cite{can,Kotovskii2015,MORGENSTERN2020}, while 
$^{227}\rm{Ac}$ is considered for use as the active element of radioisotope thermoelectric generators. In combination with beryllium, $^{227}\rm{Ac}$ is an effective neutron source \cite{Russell}, applied in neutron radiography, tomography and other radiochemical investigations. Moreover, $^{227}\rm{Ac}$ is used as a tracer for deep seawater circulation and mixing \cite{GEIBERT2002147}. On the fundamental-physics side, actinium can be considered as a possible system to study parity-nonconservation and time-reversal-invariance violation effects \cite{Dzuba2012,Roberts2013}. 
Rare isotopes of actinium are produced and were studied at different on-line facilities worldwide. These research activities started at TRIUMF, Canada \cite{sebatian2013source} and, together with contributions from the LISOL facility in Belgium \cite{Verstraelen2019}, are still ongoing. At ISOLDE CERN, production of the isotope $^{229}$Ac was investigated, acting as mother for the $^{229}$Th isomer proposed as a nuclear clock \cite{Verlinde2019}. Further rare isotopes will become available
with high yield at the Facility for Rare Isotope Beams (FRIB) \cite{Abel_2019}. Studies of rare actinium isotopes contribute to deriving nuclear physics properties and trends in this region of the nuclear chart and help to decode astrophysical processes, to understand fundamental interactions, and to develop practical applications, for example, in nuclear medicine and material sciences.
The atomic structure of actinium was elucidated by Judd who calculated the ordering and properties of low-lying levels of actinide atoms \cite{Judd}. This work was extended by calculations of energy differences between the lowest states \cite{Nugent} and 
a prediction of the parameters of electric-dipole (E1) transitions in actinium \cite{ElectricDipoleTransition} using the Hartree-Fock method, as well as other theoretical studies \cite{Brewer,transitionEnergy,Quinet,EnergyLevel,EliavEphraim}.

On the experimental side, Meggers observed 32 lines of neutral actinium in emission spectra \cite{meggers1957spectrochemistry}. Studies of Rydberg and autoionization (AI) states \cite{sebatian2013source} and high-resolution spectroscopy for  hyperfine-structure determination \cite{Sebastain2016,ActiniumNatureCom} were performed recently. The spectroscopic results elucidated fundamental nuclear structure properties, making actinium the heaviest element for which the $N=126$ shell closure was studied by laser spectroscopy \cite{ActiniumNatureCom}.
A continuation then also enabled the study of heavier actinium isotopes which are expected to have an exceptional octupole deformation \cite{Verstraelen2019} of interest for the studies of fundamental symmetries. Laser cooling and trapping of actinium is challenging due to the high complexity of its level structure and the scarcity of experimental information. 

Recently, Dzuba, Flambaum, and Roberts calculated 
atomic parameters of 86 low-lying states of neutral actinium with energies below $36\,218\,\rm{cm}^{-1}$ \cite{Dzuba}. 
Of these, only 28 levels had been confirmed experimentally prior to the present work. In particular, missing were the lowest-lying odd-parity levels $7s^{2} 7p~ ^{2}P^{o}_{1/2}$ and $7s^{2} 7p~ ^{2}P^{o}_{3/2}$, which should be directly accessible by E1 transitions from the $7s^{2}6d~ ^{2}D_{3/2}$ even-parity ground state. Since these predicted strong transitions are of primary importance for spectroscopic applications (e.g., fluorescence and photoionization spectroscopy, optical pumping, cooling and trapping, etc.), experimental confirmation and determination of these states' parameters (e.g., accurate energies, lifetimes, hyperfine structure, etc.) are urgently needed. In this work, a new theoretical calculation of actinium levels is presented which allows the determination of several atomic level properties. Therefore, the present work also sets a benchmark of theoretical accuracy in Ac, tests methods
to estimate theoretical uncertainties, and identifies future directions of theory development. Precise atomic calculations of Ac hyperfine constants and isotope shifts will be used for accurate extraction of nuclear properties from forthcoming laser-spectroscopy experiments.
\begin{figure}[t]
	\centering
	\includegraphics[width=1\linewidth]{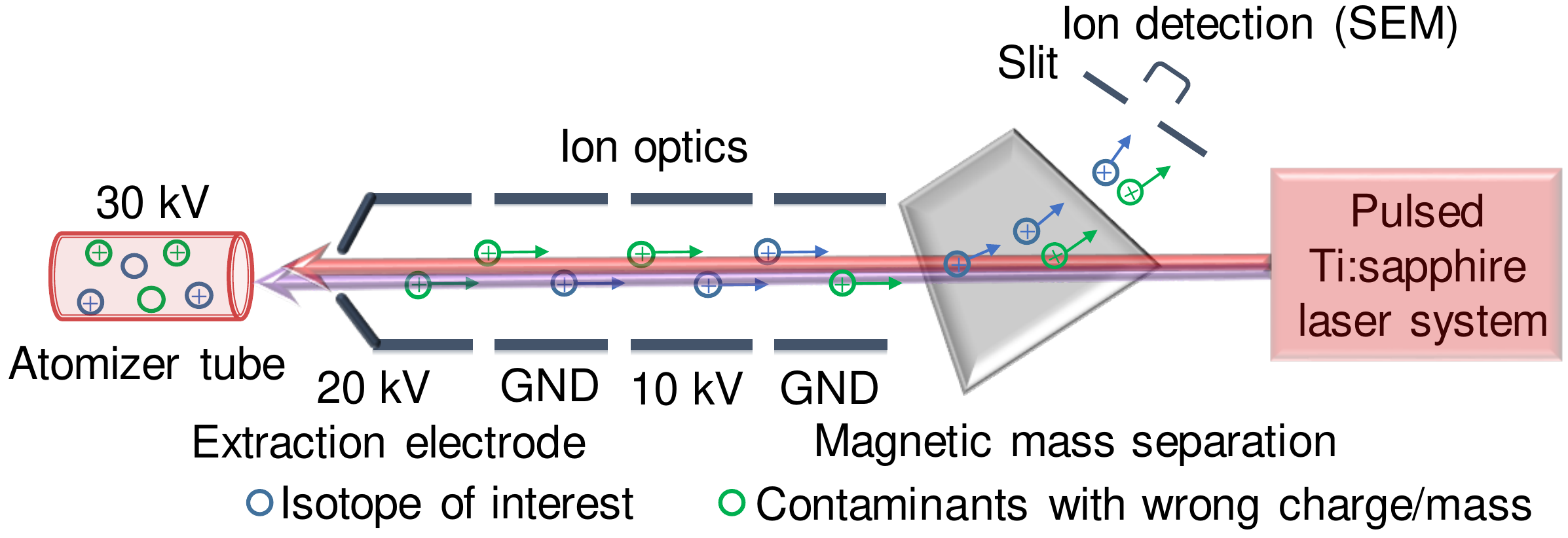}
	\caption{Overview of the off-line radioactive ion-beam facility RISIKO. Atoms are vaporized in a hot atomizer tube, where they are ionized with incident laser beams. The ions are extracted from the source region through an extraction electrode, shaped into an ion beam and guided through the magnetic mass separator. }
	\label{Fig:Setup}
\end{figure}

In this work, we used two-step, one-step resonant photoionization spectroscopy to find the two lowest odd-parity states, thus verifying the predictions of \cite{Dzuba}, precisely locating the states' energies, and determining some of their relevant properties.  
The experiments were carried out at the off-line mass separator RISIKO (resonance ionization spectroscopy in colinear geometry)
at the University of Mainz \cite{KIECK2019}. Here, a $^{227}$Ac sample of about $10^{11}$ atoms was enclosed in zirconium foil acting as reducing agent and loaded into the atomizer tube of the laser ion source (see Fig.\,\ref{Fig:Setup}). This tantalum tube is 35\,$\rm{mm}$ long with an inner diameter of 2.5\,$\rm{mm}$ and wall thickness of 1\,$\rm{mm}$, is heated up to a temperature of about 1400\,$^{\circ}\rm{C}$ to gradually atomize the sample. Two laser beams from Ti:sapphire lasers with spot size matched to the tube diameter of are directed into the source, where the actinium atoms are resonantly ionized. The ions are extracted from the atomizer tube with an extraction system with a total acceleration voltage of 30\,$\rm{kV}$. Ion optics are utilized to shape the ion beam. The accelerated ions are mass separated with a $60^{\circ}$ sector magnet and collimating slits with a resulting resolution of  $\rm{M}/(\Delta\rm{M})\approx600$. The transmitted ions are detected with a secondary electron multiplier (SEM).

The laser system used for this study is a set of pulsed Ti:sapphire lasers  with standard Z-shaped cavities \cite{Zcavity} developed at the University of Mainz. Each Ti:sapphire laser is pumped with a commercial, Q-switched frequency doubled Nd:YAG laser (532\,$\rm{nm}$) with a 10\,$\rm{kHz}$ repetition rate. The Ti:sapphire lasers produce up to  5\,$\rm{W}$ average output power; the pulse length is 40-60 $\rm{ns}$ and the spectral linewidth is typically 5-8\,$\rm{GHz}$. The laser wavelength can be set in the range from 690\,$\rm{nm}$ to 960\,$\rm{nm}$ with a birefringent filter and an etalon. Alternatively, frequency selection with a diffraction grating in Littrow configuration allows continuous tuning over the complete wavelength range with the drawback of lower output power. The accessible wavelength range could be extended with second- or third- and difference-frequency generation using a set of beta-barium-borate (BBO) crystals \cite{Sonnenschein2015}, which were angle-tuned for phase matching. The fundamental frequency and temporal structure of the laser outputs are monitored with wavelength meters (High Finesse WS6-600 and WSU-30) and fast photodiodes.

\begin{figure}[t]
	\centering
	\includegraphics[width=0.97\linewidth]{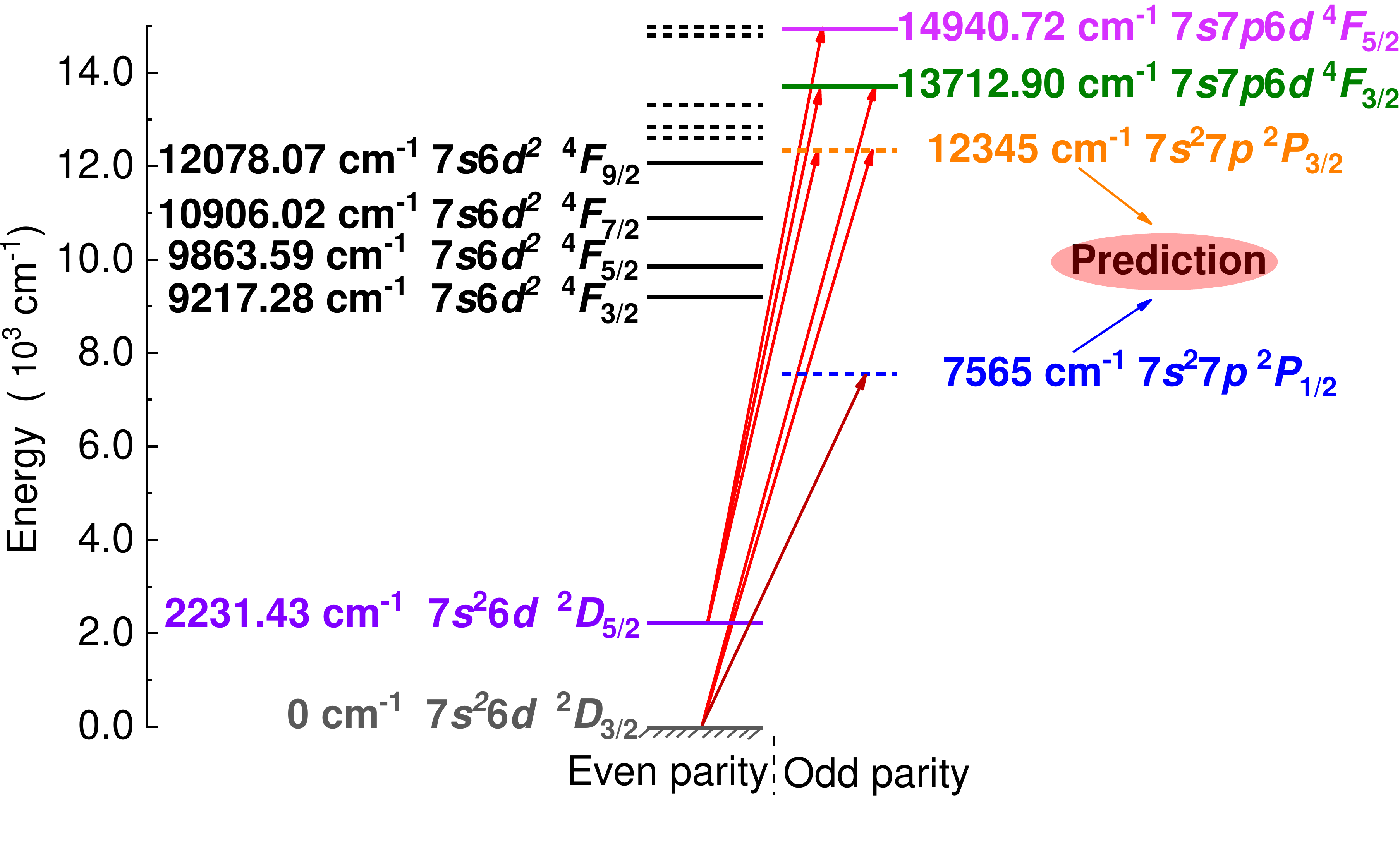}
	\caption{Low-energy levels and transitions of actinium. The predicted energies are from \cite{Dzuba}, while the other energies and assignments are from \cite{NISTdatabase}. The first-step transitions of two-step photoionization are indicated with red arrows. The colors are used to facilitate identification of the transitions in Fig.\,\ref{Fig:LongScan}.}
	\label{Fig:EnergyLevels}
\end{figure}

In our experiments, we used a two-step photoionization processes. The actinium atoms were resonantly excited to an intermediate state via a first-step transition, followed by non-resonant ionization into continuum beyond the ionization potential (IP) of $43\,394.45\,\rm{cm}^{-1}$ \cite{IPactinium}. To search for the low-lying odd-parity states predicted in \cite{Dzuba} (see Fig.\,\ref{Fig:EnergyLevels}), the frequency of the first-step laser was scanned while monitoring the ion rate. For the lower odd-parity $^{2}P^{o}_{1/2}$
state, the required infrared light was produced by difference-frequency generation (DFG) employing the 457\,nm
intracavity frequency doubled light of a standard Z-shaped Ti:sapphire laser
with a power of 1.1\,W (pump) and the fundamental output around 700\,nm, 1\,W power, from another, grating-tuned Ti:sapphire laser (idler). BBO crystals were used for both frequency doubling and DFG (type I phase matching with an angle of $24.6^{\circ}$).
The DFG setup produced about 4\,mW of light power around 1321\,nm. 
To excite the 
$^{2}P^{o}_{3/2}$ state, the light at $\approx 810\,$nm was generated with a Ti:sapphire laser. For the ionization step, we used an external third-harmonic generation of the fundamental emission of a standard Ti:sapphire laser to produce the light ($\lambda \approx$\,274\,nm; $\approx\,$85\,mW of power). Several resonances of the first excitation step were observed, originating from either the ground state or the thermally excited low-lying state at 2231.43\,cm$^{-1}$ ($7s^{2}6d\,\,^2D_{5/2}$, with a population of about 20\% of the ground-state population).

\begin{figure}[t]
	\centering
	\includegraphics[width=1\linewidth]{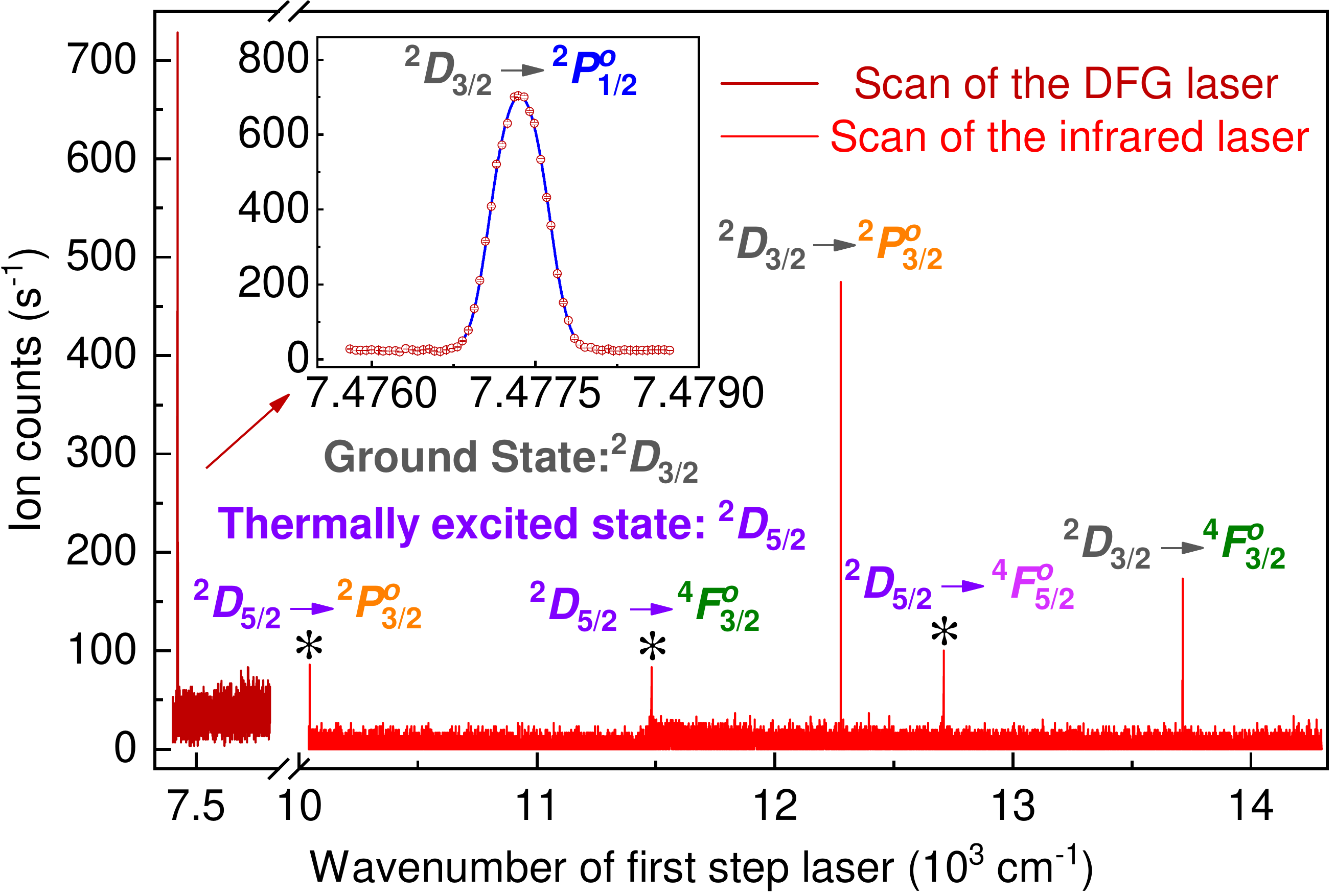}
	\caption{Observed spectra with a broad scan of the two individual first-step lasers around 7\,400 and 10\,000-14\,000\,cm$^{-1}$, respectively. Note the cut in the horizontal axis. The inset depicts a detailed scan near the transition from the ground state to the previously missing odd-parity $^2P^{o}_{1/2}$ state and a corresponding fit.}
	\label{Fig:LongScan}
\end{figure}

The results of the scans over the accessible energy ranges for the two different first-step lasers are compiled in Fig.\,\ref{Fig:LongScan}. The spectrum reveals six lines, which can be identified as the transitions shown with red arrows in Fig.\,\ref{Fig:EnergyLevels}. Three of them, marked with asterisks in Fig.\,\ref{Fig:LongScan}, are identified as transitions from the thermally populated $^2D_{5/2}$ state.  We determine the energies of the previously undetected $7s^{2}7p~^{2}P^{o}_{1/2}$ and $7s^{2}7p~^2P^{o}_{3/2}$ states as 7477.36(1)$_{\rm{stat}}(4)_{\rm{sys}}\,\rm{cm}^{-1}$ and 12\,276.59(1)$_{\rm{stat}}(2)_{\rm{sys}} \rm{cm}^{-1}$, respectively. These experimental energies are $87.6\,\rm{cm}^{-1}$ and $68.4\,\rm{cm}^{-1}$ lower than the predicted values \cite{Dzuba}, respectively. The statistical error is inferred from standard deviation, accounts for the statistical readout error of the wavenumber and the uncertainty of the fitted position. A minor contribution is ascribed to an imperfect synchronization in the data acquisition process, depending on the scan direction and speed. The systematic error is that of the wavelength measurement using the wavemeters. It was verified
by analyzing the energy positions of the $7s7p 6d\,^4F^{o}_{3/2}$, $^4F^{o}_{5/2}$ levels, which were found to be in agreement with the values in the NIST database \cite{NISTdatabase}.

\begin{figure*}[t]
	\centering
	\includegraphics[width=0.99\linewidth]{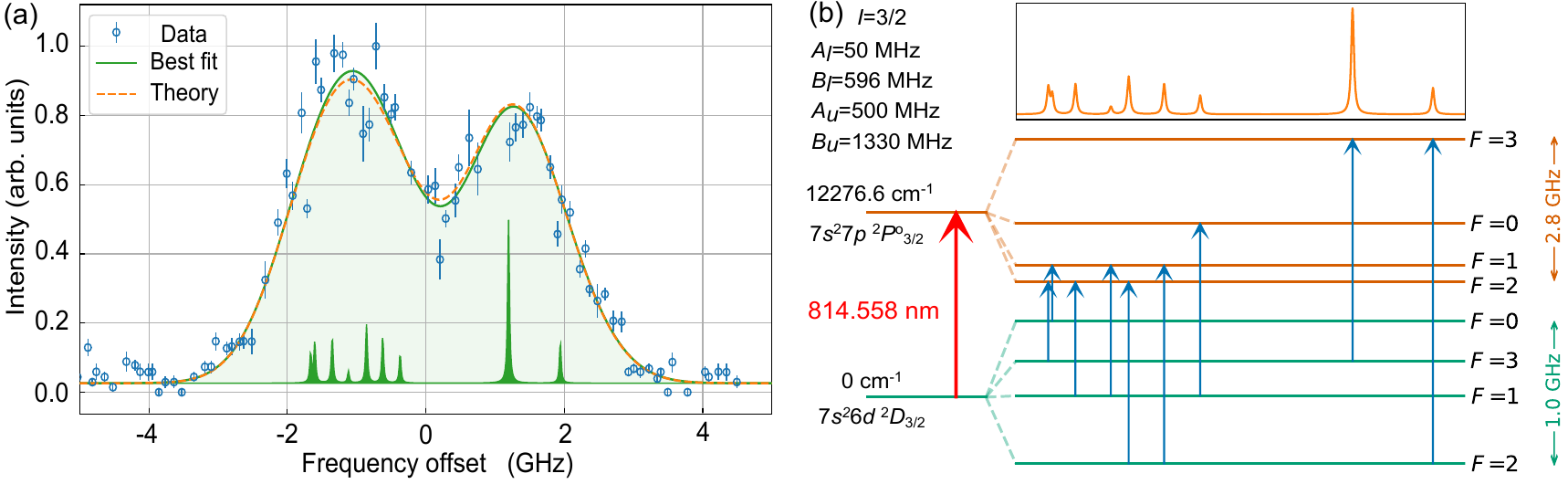}
\caption{Hyperfine structure of the transition from the ground state to the newly observed $7s^27p\ ^{2}P^{o}_{3/2}$ level. (a) The measured data are the points with error bars. The corresponding fit and theoretical profiles are shown with the green line with light-green shading under it and the dashed-orange line; the individual hyperfine components of the transition (best fit) are shown as the dark-green shaded structure at the bottom. The zero on the horizontal axis is the line centroid corresponding to 12\,276.59\,$\rm{cm}^{-1}$.  (b) Hyperfine sublevels of the ground and the odd-parity $^2P^{o}_{3/2}$ states. The hyperfine components of the transition are depicted with arrows; the predicted strengths of these components are shown at the top. 
The lower-state hyperfine constants are taken from \cite{sonnenschein2014laser} and the upper-state values were calculated in this work. The wavelength is given in vacuum.}
	\label{Fig:hyperfineStructure}
\end{figure*}

In addition to the broad scan, detailed scans near the resonances were performed, such as that shown in the inset of Fig.\,\ref{Fig:LongScan}. The crimson circles are the experimental data, which are found to be well described by a saturated Gaussian function given by
\begin{align}
	y&=y_{0}+A_{0}\frac{sG(E_{0},\sigma)}{1+sG(E_{0},\sigma)}
\ .
	\label{fitfunction}
\end{align}
Here $y_{0}$ is the background ion signal, $A_{0}$ is the amplitude of the peak, and $E_{0}$ and $\sigma$ are its centroid and linewidth of the Gaussian $G$; $s$ is the saturation parameter.  
The linewidth was determined from the fit to be $\sigma \approx 0.18\,\rm{cm}^{-1}$ (or, equivalently, 5.4\,GHz) dominated by the laser linewidth. It was possible to reduce this linewidth and perform more detailed scans of the resonances as shown in Fig.\,\ref{Fig:hyperfineStructure}(a).  To this end, the second-step ionization pulse was delayed by 80\,$\rm{ns}$ from the optimal delay to reduce line broadening due to the presence of the ionizing-laser field \cite{PhysRevA.Narrowlinewidth}, while the first-step laser was operated with an additional intracavity etalon (uncoated YAG, R=0.08, 6\,mm thickness) \cite{sonnenscheinDualEtlon}. The average power of this laser was reduced to 10\,$\rm{mW}$ to reduce saturation broadening. 

\begin{figure}[t]
	\centering
	\includegraphics[width=1\linewidth]{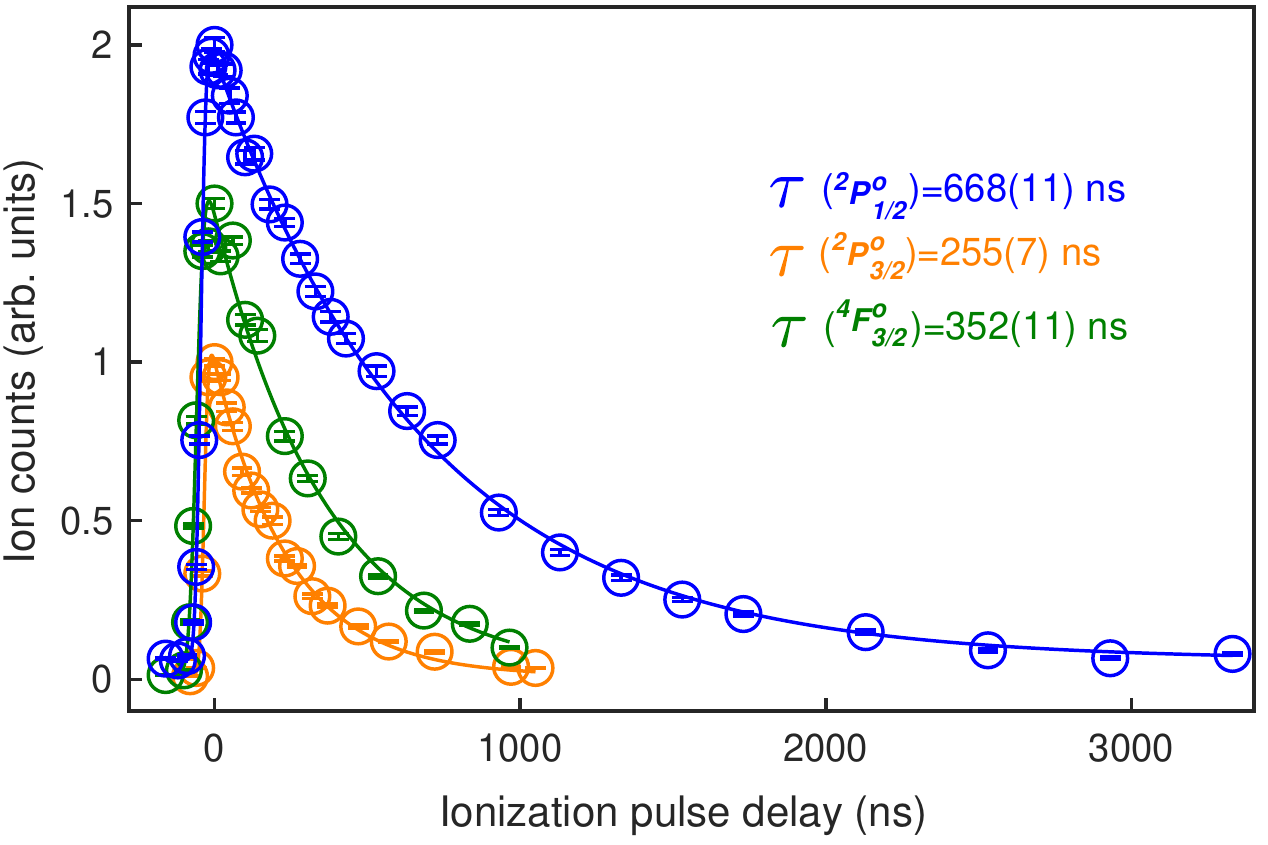}
	\caption{Lifetime measurements of the excited states. The ion signal is fitted with a convolution of a Gaussian function and exponential decay (see text). The ion counts of the individual curves feature an arbitrary scaling factor for better visualization.}
	\label{Fig:lifetime}
\end{figure}

The hyperfine sublevels of the ground and the $^2P^{o}_{3/2}$ state are depicted in Fig. \ref{Fig:hyperfineStructure} (b), involving nine hyperfine components of the transition shown with arrows. The hyperfine structure is, to a good approximation, described by the magnetic dipole and electric quadrupole coupling constants $A$ and $B$. We use subscripts ``$l$'' and ``$u$'' to indicate the hyperfine constants of the ground and  excited states, respectively. The relative intensities $S_{FF^{\prime}}$ of the transitions between individual hyperfine levels $F$ and $F^{\prime}$ with electronic angular momenta $J$, $J^{\prime}$ are related to the intensity of the underlying fine-structure transition $S_{JJ^{\prime}}$  according to (see, for example, \cite{auzinsh2010optically}):
    \begin{align}
    	\frac{S_{FF^{\prime}}}{S_{JJ^{\prime}}}=(2F+1)(2F^{\prime}+1)\left\{\begin{array}{l}
    	F^{\prime}~F^{\prime}~1 \\
    	J^{\prime}~J~~I
    	\end{array}\right\}^2\,,
    \end{align}
where the curly brackets indicate the Wigner $6j$ symbol. The open circles in Fig.\,\ref{Fig:hyperfineStructure}(a) denote the experimental data. The green line is the best-fit curve obtained with the SATLAS python package \cite{SALAS}.
The lower-state hyperfine constants were fixed at the values of $A_{l}=50.5(10)\,\rm{MHz}$, $B_{l}=596(6)\,\rm{MHz}$ taken from \cite{sonnenschein2014laser}, 
while the upper-state values as shown in Fig.\,\ref{Fig:hyperfineStructure}(b) were calculated through a hybrid approach that combines configuration interaction (CI) with a linearized coupled-cluster method that includes single and double excitations (CI+LCCSD) \cite{MS1,MS2}. Using 
the nuclear magnetic-dipole moment $\mu=1.07(18)\mu_N$, where $\mu_N$ is the nuclear magneton, and the spectroscopic electric-quadrupole moment  $Q = 1.74(10)$~eb  for the $^{227}$Ac \cite{ActiniumNatureCom}, we obtain  $A_{u}=499(21)(84)\,\rm{MHz}$, $B_{u}=1332(130)(77) \,\rm{MHz}$, where the first number in parenthesis is the estimated uncertainty of theoretical calculations, and the second number comes from the uncertainty in the values of the nuclear moments. These values were used to calculate the expected structure shown in Fig.\,\ref{Fig:hyperfineStructure}(b). 
Using these as starting parameters and fixing the well-known ground-state constants, a best fit to the data [the green curve in Fig.\,\ref{Fig:hyperfineStructure}(a)] results in $A_{u}=513(8)\,\rm{MHz}$, $B_{u}=1260(52)\,\rm{MHz}$. These values show an excellent agreement with the prediction [the dashed orange curve in Fig.\,\ref{Fig:hyperfineStructure}(a)], also confirming the present nuclear magnetic-dipole moment value from  \cite{ActiniumNatureCom}. Since the theoretical uncertainty for $A(^2P_{3/2})$ is found to be 4\%, it is particularly
suitable for significantly improved extraction of the nuclear magnetic moment.  Note that the present hyperfine-structure results also agree with earlier, lower-precision, theoretical predictions for the upper state by Beerwerth \cite{hyperfineConstant}.



The experimentally observed total linewidth (the FWHM of best-fitting Gaussian profile) of 1580(30)\,MHz is dominated by the laser width of about 1.4\,GHz. It also includes Doppler broadening of about 720\,$\rm{MHz}$ for a transition wavelength of 815\,$\rm{nm}$ and a temperature in the atomizer tube of $T\approx1400~^{\circ}\rm{C}$.
The high-resolution spectra at the bottom of Fig.\,\ref{Fig:hyperfineStructure}(a) (best fit) and top of Fig.\,\ref{Fig:hyperfineStructure}(b) (predicted from theory) indicate the positions and strengths of the individual hyperfine components assuming an artificial linewidth fixed at 30\,MHz for visualization. A high-resolution study of the hyperfine structure of the excited atomic states will be a topic of future research.

For further characterization, the lifetimes of the excited $^2P^{o}_{1/2}$, $^2P^{o}_{3/2}$, $^4F^{o}_{3/2}$ states were measured using a delayed ionization technique. Following the first-step laser pulse, the second-step ionization laser pulse was applied after a variable delay. Figure \ref{Fig:lifetime} shows the excited-state population decay as a function of the ionization-pulse delay. The evolution of the population during the ``dark'' time between the pulses corresponds to an exponential decay. To be able to also include the initial part of the profile overlapping with the first-step laser profile, the influence of the laser field was modelled by assuming a convolution of a Gaussian profile and an exponential decay \cite{king1975lifetimeFitting}. We note that the present lifetime-determination method is not applicable to lifetimes much shorter than the laser-pulse duration ($\approx 50$\,ns) and much longer than the collisional lifetime of the excited atoms within the laser beam in the atomizer tube ($\approx3~\rm{\mu s}$) \cite{PhysRevA3us.98.042504}. The experimentally observed lifetimes listed in Table \ref{tab:Summary} are safely within this range.

\begin{table*}
	\setlength\tabcolsep{0pt}  
	\setlength\extrarowheight{1mm}
	\centering
	\caption{The determined excitation energies and lifetimes, and comparison with theory and literature.}
	\begin{tabular*}{\textwidth}{ S[table-format=3.2] @{\extracolsep{\fill}} *{9}{c} }
		\hline
		\hline\\[-0.3cm]
		\multicolumn{2}{c}{State} &
		\multicolumn{3}{c}{Energy\,($\rm{cm}^{-1}$)} &
		\multicolumn{4}{c}{$\tau~(\rm{ns})$}  \\
		 \cline{3-5}   \cline{6-9}
		 &  &
		 Exp. & Calc. & Lit. &  Exp. & Calc.\,I & Calc.\,II & Lit. \cite{Dzuba,Dzuba2020}  \\
		\hline
		 $ {7s^27p} $ & $^2P^{o}_{1/2}$ & $7477.36(4)$ & $7701(250)$ & $7565$ \cite{Dzuba} &  $668(11)$& $647$ & $707(53)$ & $733(70)$ \\
		 ${7s^{2}7p}$ & $^2P^{o}_{3/2}$ & $12\,276.59(2)$ & $12\,475(250)$ & $12\,345$\,\cite{Dzuba}& $255(7)$&$209$& $219(16)$&$238(20)$\\
		 ${7s7p6d}$ & $^4F^{o}_{3/2}$ & $13\,712.74(3)$ &  $13\,994(370)$& $13\,712.90$\,\cite{NISTdatabase}&$352(11)$& $327$& $351(29)\,$ &$317(30)$\\
		\hline
	\end{tabular*}
\label{tab:Summary}
\end{table*}

A comparison of the calculated  and experimentally determined energies and lifetimes is shown in Table \ref{tab:Summary}. While we list results for the three states of experimental interest, we calculated energies of 18 states using the CI+LCCSD approach, including 114\,840 configurations and demonstrated convergence of the results with the increasing number of configurations. QED and full Breit corrections are included as described in Refs.\,\cite{MS3,MS4}. Our results for even and odd levels agree with previous  experiments \cite{NISTdatabase} to 40-120 cm$^{-1}$ and 200-350~cm$^{-1}$, respectively, with theory values being larger than the experimental ones in all cases.  Such regular differences with experiment let us predict that we overestimate the energies of the $^2P_J$ levels by about 200~cm$^{-1}$, with about 50~cm$^{-1}$ uncertainties which is in excellent agreement with measured values.
We list the lifetimes obtained using theoretical values of energies and electric-dipole (E1) matrix elements in the column labeled ``Calc.\,I''. We use experimental energies and theoretical values of  E1 matrix elements to calculate the final theoretical lifetimes listed in column labeled ``Calc.\,II''. The uncertainties in the lifetimes are estimated from the size of the 
higher-order corrections to E1 matrix elements  determined from the difference of the CI+LCCSD values and another calculation that combines CI with many-body perturbation theory \cite{MS5}.

Note that the lifetime values listed in \cite{Dzuba} were recently corrected \cite{Dzuba2020}; the corrected values are given in Table \ref{tab:Summary}. 
Within the respective uncertainties, there is agreement between the two independent calculations and the experiment. 
 
In summary, using the two-step, one-step resonant photoionization, we have located the two lowest-lying odd-parity states in Ac as predicted by theory \cite{Dzuba}. We have measured the energies and lifetimes, as well as hyperfine parameters of the $^{2}P^{o}_{3/2}$ state, once again, in agreements with theoretical predictions and implying good understanding of the atomic structure of the actinium atom. These findings will aid in developing techniques for cooling and trapping of actinium,  as well as in optimization of specific resonance-ionization processes.  
The results will be useful for production of $^{225}$Ac for nuclear medicine, and may support the design of fundamental-physics experiments such as investigations of fundamental symmetries with this atom. In addition, they provide a test and validation of the advanced many-body atomic theory making a foray into high-precision calculations of the highly complex actinide spectra. 

\ 

The authors thank V.\,V. Flambaum and V.\,A. Dzuba for stimulating discussions and providing the corrected lifetimes, and M. Block, Ch. Mokry and J. Runke for providing the actinium sample. We gratefully acknowledge discussions with R. Beerwerth and S. Fritzsche on the theory of the actinium hyperfine structure. The work was supported by the German Federal Ministry of Education and Research under the project numbers 05P15UMCIA, 05P18UMCIA. DB was supported in part by the DFG Project ID 390831469:  EXC 2118 (PRISMA+ Cluster of Excellence). DB also received support from the European Research Council (ERC) under the European Union Horizon 2020 Research and Innovation Program (grant agreement No. 695405), from the DFG Reinhart Koselleck Project. Theory work was supported in part by  U.S. NSF Grant No.\ PHY-1620687. SGP and MGK acknowledge support by the Russian
Science Foundation under Grant No.~19-12-00157.

\bibliographystyle{apsrev4-1}
\bibliography{DetectionNewAtomicState.bib}

\end{document}